\documentstyle[12pt]{article}

\begin{document}

%\mbox{ }\\
%\hfill
%{\large\bf AZPH-TH/94-20}\\
%\preprint{AZPH-TH/94-20}

\begin{center}
{\bf Charm Production in Hadronic and Heavy-Ion Collisions\\
at RHIC and LHC Energies to $O(\alpha_{s}^3)$}
\end{center}

\begin{center}
Ina Sarcevic and Peter Valerio\\
Department of Physics, University of Arizona, Tucson AZ 85721
\end{center}
\vskip 1true in

\begin{abstract}
We present results on rapidity and transverse momentum
distributions of inclusive charm quark production in
hadronic and heavy-ion collisions at RHIC and LHC energies,
including the next-to-leading
order, $O(\alpha_s^3)$, radiative corrections and the
nuclear shadowing effect.
We determine the hadronic and the {\it effective} (in-medium)
K-factor
for the differential and total inclusive charm cross sections.
We find the
fraction of central and inelastic events that contain at
least one charm
quark pair at LHC energies and
obtain the effective $A$-dependence of the inclusive
charm production in
proton-nucleus and nucleus-nucleus
collisions at RHIC and in nucleus-nucleus collisions at the LHC.
We discuss theoretical uncertainties inherent in our calculation.
In particular, we show how
different extrapolations of
gluon density
in a nucleon {\em and} in a nucleus to the low-$x$ region
introduce large theoretical uncertainty in the calculation of
charm production at LHC energies.
\end{abstract}
\vfil\eject

\noindent{\bf 1.  Introduction}
\medskip

Recently, there is a considerable theoretical and experimental
interest in studying charm production in hadronic and
nuclear collisions.  Theoretical calculation of the
heavy-quark differential and total cross sections has been
improved by including the next-to-leading
order, $O(\alpha_s^3)$,
radiative corrections [1,2].  For bottom and charm production
these corrections are large, especially at threshold energies and
at very high energies.
Future high-precision measurements of the charm
production at Fermilab fixed-target experiments
could therefore provide a
stringent test of perturbative QCD [3].
In addition,
studying charm production at the proposed
heavy-ion colliders, BNL's Relativistic Heavy Ion
Collider (RHIC) and CERN's Large Hadron Collider (LHC) is of
special interest.
Open charm production has
been suggested as an elegant signal for detecting the
formation of quark gluon plasma (QGP) in heavy-ion collisions
[4].
If thermalization is reached
in heavy-ion
collisions at RHIC
%($\sqrt{s}\approx 200$GeV/$A$)
and LHC,
%($\sqrt{s}\approx 7$TeV/$A$)% energies are sufficient
we expect a very dense matter, of the order of
few
GeV/$fm^{3}$, to be formed in the
initial stage of the collision.  It seems plausible that this
density is sufficient for creating a new
state of matter, the QGP [5].
The
search for a clean, detectable signal for this new state of matter
has been one of the most challenging
theoretical problem for the last few years.  Some of the
proposed signals, thermal photons, dileptons and $J/\Psi$
suppression, have been studied and found to be difficult to detect due
to the large QCD
background [6].
In
order that the enhanced charm production can be used as the signal of
QGP, we need to understand the
QCD background, namely
the production of charm quarks through the
hard collisions of partons inside the nuclei.
This type of charm production
at RHIC and LHC energies is dominated
by initial-state gluons.  Therefore, in addition to the
possibility of
pointing towards the formation of quark-gluon plasma in
high-energy heavy-ion collisions,
combined measurements of charm
production in p-p, p-A and A-A collisions could
provide valuable information about the
gluon density in a nucleus.

In section II we present our calculation of the
total cross section for open charm production
in hadronic collisions and compare our results
with the low energy
measurements.  Our
calculation includes the next-to-leading order,
$O(\alpha_s^3)$,
radiative corrections.
We discuss theoretical uncertainties
due to the choice of the factorization and renormalization scale and
the choice of the parton structure function.  We show how the low
energy measurements of the total cross section for charm
production in p-p and
p-A collisions provide certain constraints on our theoretical
parameters.
We illustrate the importance of
including the
next-to leading order
corrections at these energies.
We present our predictions for
the total cross section for
charm production
in p-p collisions at RHIC and LHC energies.
In Section III we review nuclear effects of relevance to charm
production in nuclear collisions.  In particular, we
obtain the number of nucleon-nucleon collisions per unit
transverse area at fixed impact parameter,
the so-called spatial overlapping function,
for two different choices of the nuclear density, the
Woods-Saxon [7] and the Gaussian form.  We also discuss the
nuclear shadowing effect in the standard parton model and
recent direct measurements of this effect on
quark distribution in a nucleus.
We find the effective A-dependence of the total
cross section for charm production in A-A collisions at
RHIC and LHC and determine the fraction of central and inelastic
events that contains
at least one charm quark pair at these energies.
In section IV we present
rapidity and transverse momentum distributions for
charm production in hadronic and nuclear collisions.
We illustrate the importance of the $O(\alpha_s^3)$
corrections
{\it combined} with nuclear shadowing effects
in different regions of phase space by
calculating the
{\em effective} (i.e. in-medium)
K-factor, defined as the ratio of the inclusive distribution for
charm production in A-A collisions to the
leading-order distribution without nuclear effects.
We show that the K-factor, in general, is a
function of rapidity,
transverse
momentum and $x_F$ ($x_F=p_z/p_{z max}$, and $p_z$ is taken to be
in the beam direction) and can not be taken to be a constant.
In section V we present our conclusions.

\medskip
\noindent{\bf 2.  The Total Cross Section for Charm Production in
Hadronic Collisions}
\medskip

In perturbative QCD, the total inclusive cross section
for charm production in hadronic collisions is obtained
as a convolution of parton densities in the hadron
with the hard
scattering cross section.  Our calculation
includes both the leading order subprocesses, $O(\alpha_s^2)$, such as
$ q+\bar q\rightarrow Q+\bar Q$ and
$ g+g\rightarrow Q+\bar Q$,
and next-to-leading order subprocesses, $O(\alpha_s^3)$, such as
$ q+\bar q\rightarrow Q+\bar Q +g$,
$ g+q \rightarrow Q+\bar Q+g$,
$ g+\bar q \rightarrow Q+\bar Q+\bar q$ and
$ g+g \rightarrow Q+\bar Q+g $.   The total inclusive cross section
for charm production in hadronic collisions can be written as
\begin{equation}
\sigma_c=\int_{\frac{4m^{2}_{c}}{s}}^{1}dx_{a}\int_{\frac{4m^{2}_{c}}
{x_{a}s}}^{1}dx_{b}\sum_{i,j}^{partons}[F_{i}(x_{a},Q^{2})
F_{j}(x_{b},Q^{2})\hat{\sigma}_{i,j}(\hat{s},m_{c}^{2},Q^{2})],
\end{equation}
where the $F_{i}(x,Q^{2})$'s are the parton distributions in a nucleon,
$x_a$ and $x_b$ are the fractional momenta of incoming partons and
$\hat {s}= x_a x_b s$ is the parton-parton center of mass energy.
The parton cross section $\hat\sigma_{i,j}(\hat s, m_c^2, Q^2)$
has been calculated to the order O($\alpha_s^3$) and can be written
as
\begin{equation}
\hat \sigma_{i,j}(\hat s, m_c^2, Q^2)={\alpha_s^2(Q^2)\over
{m_c^2}}
f_{i,j}(\rho, {Q^2\over{m_c^2}}),
\end{equation}
where
\begin{equation}
f_{i,j}(\rho,{Q^2/{m_c^2}})=f_{i,j}^{(0)}(\rho)+4\pi\alpha_s(Q^2)
[f_{i,j}^{(1)}(\rho)+\bar f_{i,j}^{(1)}(\rho)\ln {({Q^2/{m_c^2}})}
].\end{equation}
The functions $f_{i,j}^{(0)}$,$f_{i,j}^{(1)}$ and
${\bar f}_{i,j}^{(1)}$  are given in Ref. 1.  The coupling  constant
$\alpha_{s}(Q^{2})$, calculated to next-to-leading order,
is given by
\begin{equation}
\alpha_s (Q^2)={12\pi\over
{(33-2N_f)\ln {Q^2\over{\Lambda^2}}}}[1-{6
(153-19N_f){\ln \ln {Q^2/
{\Lambda^2}}}\over{{(33-2N_f)}^2\ln {Q^2/\Lambda^2}}}],
\end{equation}
where $Q^2$ is the renormalization scale, $\Lambda$ is the QCD scale
parameter and $N_f$ is the number of flavors.
%% discussion of scales and dependence on the scale.
We take the factorization scale in the structure functions to be
$2m_c$ and we consider the renormalization scales
$Q=m_c$ and $Q=2m_c$.
For the mass of the charm quark we use
$m_c=1.5$GeV.
We do not consider the factorization scale below
$2m_c$ because the structure functions have been measured only  for
$Q^2\geq 8.5$GeV [8].
%%discussion of structure functions
In our calculation we use two-loop-evolved parton
structure functions:
\par
1)  Martin, Roberts and Stirling, MRS S0 [9], with
$\lambda_{5}=140$ MeV.
\par
2)  Martin, Roberts and Stirling, MRS D0 [9], with
$\lambda_{5}=140$ MeV.
\par
3)  Martin, Roberts and Stirling, MRS D-- [9], with
$\lambda_{5}=140$ MeV and ``singular'' behavior of the
gluon distribution at small $x$, i.e. $G(x,Q^2)\sim x^{-1.5}$.
\par
4)  Martin, Roberts and Stirling, MRS A [10], with
$\lambda_{5}=151$ MeV and ``singular'' behavior of the
gluon distribution at small $x$,
i.e. $G(x,Q^2)\sim x^{-1.08}$.  This is the most recent MRS set,
adjusted to fit recent
H1 and ZEUS data [8].

In Fig. 1 we present our
results
for the total cross
section for charm production in proton-proton collisions for
the beam energies ranging from 50GeV up to 2TeV.
The results were
obtained using two different sets of structure functions, MRS D0
(dashed lines) and MRS A (solid lines).  We find the
uncertainty due to the choice of  structure function to be
only few percent.  This is not surprising, because the range of
$x$ probed by
charm quark pair production at these energies is well
within the range of the
data for nucleon structure functions.
The top two curves in Fig. 1
correspond to the
calculation using
the renormalization scale
$Q^2=m_{c}^{2}$ and the bottom two to
$Q^{2}=4m_{c}^{2}$.  Comparison of our results
with low energy
data for
p-p and p-A collisions [11] indicate better agreement
for the choice of
the renormalization scale
$Q^{2}=m_{c}^{2}$.
Future high-precision charm experiments at
Fermilab [3] might be able to provide a tighter constraint on this
theoretical parameter.
\par
We find that the contribution
from higher-order corrections,
($O(\alpha_s^3)$),
are important
at low energies, especially near the threshold energy for
charm production.  The K-factor, defined as a ratio of the
next-to-leading order result to the leading-order one,
ranges from
about $3.7$ at $E_{beam}=50$GeV to about $2.2$ at
$E_{beam}=2$TeV.

We predict that the total cross section for
charm production in p-p collisions is  $180\mu b-210\mu b$
(for $Q^2=m_c^2$) and $112\mu b-126 \mu b$ (for $Q^2=4m_c^2$)
at RHIC and $2.4 mb - 15.3mb$
(for $Q^2=m_c^2$) and $1.4 mb - 9.2 mb$
(for $Q^2=4m_c^2$)
at  LHC.
The range of the cross sections correspond to
different
structure functions.
At
RHIC energies, we find that
the cross section depends weakly
on the
choice of the structure function.
The average $x$-value that is probed with charm
production at $\sqrt{s}=200$GeV is on the order of
$10^{-2}$, still within the range of $x$ for which there is
deep
inelastic scattering data.  Theoretical
uncertainty due to the
choice of the structure function is about $15\%$.
At
LHC energies
($\sqrt{s}=7$ TeV) the average $x$-value probed with charm
production in the central rapidity region is about
$5\times10^{-6}$, far below the $x$-range covered by
the current deep
inelastic scattering data.  Different
sets of structure functions [9,10], which all fit the
current data\footnote{
The most recent H1 and ZEUS data on
structure functions
at low-$x$ and low-$Q^2$ [8] are best fitted with MRS A
distributions.
The
data seem to be steeper function of $x$ than
the MRS D0 structure functions
and not as steep as the MRS D--[9,10].}
have different extrapolation to
low-$x$  region.
For example, the
MRS D-- and MRS A gluon distribution have singular behavior
in the small-$x$ region, in contrast to
MRS D0 and MRS S0 sets.
As a consequence,
the total cross section for charm production at  LHC
calculated with MRS D-- parton
distributions is larger than the one obtained with
MRS D0 set by about a factor of $6$. We find the
K-factor for the total cross section for charm production
be between $2$ and $2.1$ at RHIC and
between $2.3$ and $2.8$ at  LHC.
The range for the K-factor corresponds to  the
choice of the renormalization  scale.

\medskip
\noindent{\bf 4.  Charm Production in Nuclear Collisions}

\medskip
\noindent{\bf a)  Nuclear Geometry}
\medskip

Assuming the validity of factorization theorems
in the calculation of the cross section for
charm production in nuclear collisions,
the total number of charm quark pairs
produced in A-A collisions
at some fixed impact parameter is given by [12]
\begin{equation}
N_c
^{AA}(s,b)
=\int d^{2}b_1
dx_{a}
dx_{b}\sum_{i,j}^{partons}[F_{i}^A(x_{a},Q^{2},
\vert\vec{b}-\vec{b_1}\vert )
F_{j}^A(x_{b},Q^{2},\vert \vec{b_1}\vert)
\hat{\sigma}_{i,j}(\hat{s},m_{c}^{2})],
\end{equation}
where the $F_{i}^{A}(x,Q^{2},b)$ is the parton distribution
function in a nucleus.
We assume that the parton density in a nucleus can be
factorized in terms of the (usual) parton structure function
modified by the medium effects,
$F_i^A(x,Q^2)/A$,
and the spatial distribution of partons at some impact
parameter $b$, $T_{A}(b)$, i.e.
\begin{equation}
F_{i}^{A}(x,Q^{2},b)\approx (F_{i}^{A}(x,Q^{2})/A) \\T_{A}(b).
\end{equation}
The nuclear thickness function, $T_{A}(b)$, is
the number of
nucleons per unit transverse area at fixed impact parameter.
The spatial (impact parameter) integration that appears in
Eq. (5)
gives the
number of
nucleon-nucleon collisions
per unit of transverse area at fixed impact
parameter, $T_{AA}(b)$, which is related to the
nuclear density in a following way [13]
\begin{equation}
T_{AA}(b)=\int d^{2}b_{1}
T_{A}(\mid\vec{b}_{1}\mid)
T_{A}(\mid\vec{b} - \vec{b}_{1}\mid),
\end{equation}
where
the nuclear thickness function,
$T_A(b)$, is the
nuclear density integrated over the longitudinal size, i.e.
\begin{equation}
T_{A}(b)=
\int_{-\infty}^{\infty}dz \rho_{A}(\sqrt{b^{2}+z^{2}}).
\end{equation}
For A-A collisions we take the nuclear density to be the
Woods-Saxon
distribution [7]  given by
\begin{equation}
\rho(r)=\frac{n_{0}}{[1+e^{{(r-R_{A})}/d}]},
\end{equation}
where $n_{0}\approx 0.17/fm^{3}$, $R_{A}$ is the nuclear radius and
$d=.54 fm$ is the ``skin" thickness of this distribution.
The density and the
nuclear overlapping function are normalized so that
$\int d^{3}  r \rho( r) = A$ and
$\int d^{2} b T_{AA}(b)=A^{2}$.
For central collisions
the
overlapping function
can be approximated
by $T_{AA}(0)={A}^{2}/\pi {R_{A}}^{2}$, which gives
$T_{Au-Au}(0)=30.7mb^{-1}$.

The nuclear density that is most widely used in the
literature is the
Gaussian
distribution
\begin{equation}
\rho(r)=\frac{A}{\pi^{3/2}a^{3}} e^{-r^2/a^2},
\end{equation}
where $a$ is related to the charge radius of the nucleus,
$\frac{3}{2}a^{2}=
\langle R_{A}^{2}\rangle$.
Functions
$T_{A}(b)$ and $T_{AA}(b)$ can be
obtained analytically and are given by
\begin{equation}
T_{A}(b)=\frac{A}{\pi a^{2}}e^{-\frac{b^{2}}{a^{2}}}
\end{equation}
and
\begin{equation}
T_{AA}(b)=\frac{A^{2}}{2\pi a^{2}}e^{-\frac{b^{2}}{2a^{2}}}.
\end{equation}

The formalism for proton-nucleus collisions is similar.  The
thickness function of the proton is approximated by
\begin{equation}
T_{p}(b)\simeq \frac{1}{(2\pi)^{2}} \int d^{2}k_{\perp}
G_{E}(k_{\perp}^{2}) e^{i{\bf k_{\perp}\cdot b}},
\end{equation}
where $G_{E}$ is the proton electric form factor
\begin{equation}
G_{E}(k_{\perp})\simeq (1+\frac{k_{\perp}^{2}}{\nu^{2}})^{-2}.
\end{equation}
The corresponding overlapping
function is given by [14]
\begin{equation}
T_{pA}(b)=\frac{\nu^{2}A}{48\pi a^{2}} \int_{0}^{\infty}db' b'
e^{-(b^{2}+b'^{2})/a^2}(\mu b')^{3} K_{3}(\mu b')
I_{0}(\frac{2bb'}{a^2}),
\end{equation}
where $\nu^2\simeq0.71 ($GeV$)^{2}$.   The expression for $T_{pA}$
derived from a Woods-Saxon distribution can not be obtained
analytically.
In Fig. 2. we present
$T_{p-Au}(b)$ and $T_{Au-Au}(b)$
obtained with
Woods-Saxon and with the Gaussian nuclear density distributions.
Although the Gaussian distribution is widely used,
the corresponding overlapping function has a long tail and
at high energies gives a total inelastic cross section
that violates unitarity (i.e. the cross section is larger
than the size of the physical system).

\medskip
\noindent{\bf b)  The Nuclear Shadowing Effect}
\medskip
\par
Calculation of the
charm production in nuclear collisions requires knowledge of
the nucleon structure function {\it in-medium}, $F_{i}^A
(x,Q^2)/A$, introduced in Eq. (6).
If
nucleons were independent parton densities
in a nucleus, $F_i^A(x,Q^2)$, would be simply given
as A times the parton density in a nucleon.
However, at high
energies, the parton densities become so large that the sea quarks and
gluons overlap
spatially and the nucleus can not be viewed as a collection of
uncorrelated
nucleons.
This happens when the longitudinal size of the parton, in
the infinite momentum frame of the nucleus, becomes larger than the
size of the nucleon.  Partons from different nucleons start
to interact and through annihilation effectively reduce the parton
density in a nucleus.  When partons inside the nucleus completely
overlap, they reach a saturation point.  Motivated by this simple
parton picture of the nuclear shadowing effect and taking into account
the $A^{1/3}$
dependence obtained by considering the modified, nonlinear
Altarelli-Parisi equations with gluon recombination included, the
modifying factor to the
parton structure function in a nucleus can be written as [15]
\begin{equation}
R(x,A)\equiv \frac {F_{i}^A(x,Q^2)}{A F_{i}^N(x,Q^2)}
=\left\{\begin{array}{ll}
1-\frac{3}{16}x+\frac{3}{80}&.2<x\leq 1\\
1&x_{n}<x\leq .2\\
1-D(A^{1/3}-1)\frac{1/x-1/x_{n}}{1/x_{A}-1/x_{n}}&x_{A}\leq x\leq x_{n}\\
1-D(A^{1/3}-1)&0<x<x_{A}
\end{array}
\right.
\end{equation}
where
$F_{i}^N(x,Q^2)$ is the parton structure function in a nucleon,
$x_n$ is the value of $x$ which specifies
the onset of the shadowing effect
($x_{n}=1/(2r_{p}m_{p})\approx 0.1$),
$x_A$ corresponds to the saturation point
($x_{A}=1/(2R_{A}m_{p})$,
$m_p$ is the mass of the proton,
$r_{p}$ is the radius of a proton and
$R_{A}$ is the
radius of the nucleus.
It is important to note that $x_n$ is fixed for {\em all} nuclei, and
$x_A$ can be determined for each nucleus.  Thus,
the only parameter that
is free to be fitted is $D$.
In Fig. 3 we plot $R(x,A)$ given by Eq. (16)
by fitting
$D$ to
deep inelastic lepton-nucleus data on the
ratio $F_2^A(x,Q^2)/F_2^D(x,Q^2)$ [16].
We
find that $R(x,A)$ has a
much steeper $x$-dependence than the data, especially
for
$0.002\leq x \leq 0.1$, the region
of relevance to charm production
at RHIC and LHC energies.
Even the best fit
overestimates the observed shadowing
effect by about
$15\%$.
Consequently
charm production
calculated with this shadowing function
would be underestimated
by about $40\%$.

At low energies, the effect from shadowing is ranging from
$3\%$ at $E_{beam}=50$GeV to $6\%$ at
$E_{beam}=2$TeV in p-Au collisions and from
$6\%$ at $E_{beam}=50$GeV to $13\%$ at
$E_{beam}=2$TeV in Au-Au collisions.  This effect is even
smaller for a lighter nucleus.  The reason that the nuclear
shadowing effect is so small is due the fact that
the $x$-region probed by charm production at these
energies is $0.05\leq x\leq 1$, not too far from
the value for the onset of shadowing, $x_n\approx 0.1$.

Recently, a new
parametrization of the nuclear shadowing function given by [17]
\begin{equation}
R(x,A)
=\left\{\begin{array}{ll}
\alpha_3 -\alpha_4 x & x_0 <x\leq 0.6 \\
(\alpha_3 -\alpha_4 x_0)\frac{1+k_q \alpha_2 ({1/x}-1/x_{0})}
{1+k_q A^{\alpha_1}({1/x}-1/x_{0})}
&x\leq x_{0}\\
\end{array}
\right.
\end{equation}
has been proposed.  This new function gives
much better description of all
EMC, NMC and E665 data [16] than
the shadowing function of
Eq. (16).  This is illustrated in Fig. 3.
The parameters $k_q$, $\alpha_{1}$,
$\alpha_2$, $\alpha_3$ and $x_0$ are fitted to
deep-inelastic
data  for the
ratio $F_2^A(x,Q^2)/F_2^D(x,Q^2)$ [16] and can be found in Ref. 17.
In our calculation of charm production in
p-Au and Au-Au collisions we use
the nuclear shadowing function given by Eq. (17).

Presently
there is no theory which
can quantitatively describe the observed
nuclear shadowing effect [16].
Recent calculations of the
perturbative gluon shadowing seem to substantially underestimate
the observed effect, indicating perhaps that the non-perturbative
effects are large and can not be neglected [18].
Better understanding of the nuclear shadowing effect
might require a novel,
non-perturbative approach.

To obtain the
effective A-dependence of the
total inclusive charm cross section in nuclear collisions,
defined as
$\sigma_{c}^{AA}=A^{\alpha_{eff}} \sigma_{c}^{pp}$, we use
the total charm cross section in hadronic collisions at
$\sqrt s=200$GeV ($\sqrt s =7$ TeV)
to be $\sigma_c^{pp}=180\mu b$
($\sigma_c^{pp}=2.4mb$).
The
corresponding
$\alpha_{eff}$ for central (inelastic) Au-Au collisions is
$1.27$ ($1.94$) at RHIC and $1.2$ ($1.87$) at the LHC.

\par
To be able to determine the fraction of central or inelastic
events which contain at least one charm quark pair, we need to
consider the
semiclassical probability
of having at least one parton-parton collision at fixed impact
parameter,
$1 - e^{-N_c(b)}$,
where
$e^{-N_c(b)}$
is the probability that there is
{\it no} parton-parton scattering in Au-Au collision at
impact parameter $b$.
The fraction of events in Au-Au
collisions that  contain at least one charm quark pair is then given by

\begin{equation}
\frac {\sigma_{c}^{AA}}{\sigma_{inelastic}^{AA}}
=\frac {\int d^{2}b[1-\exp(-N_c(b))]}
{\int d^{2}b[1-\exp(-T_{AA}(b)\sigma_{in}^{pp})]},
\end{equation}
where $N_c$ is given by Eq. (5).

To determine the fraction of
all {\it central} events that contain at least one charm quark pair we
integrate Eq. (18) over a small range of impact parameter, i.e.
$0\leq b \leq 0.1fm$.  We find that $98\%$ ($99\%$) of central events
at RHIC (LHC) energies will contain at least
one charm quark pair.
For {\it inelastic} collisions we integrate Eq. (18)
over all
impact parameter and find this fraction
to be $38\%$ at RHIC and ranging from
$54\%$ to $72\%$ at the LHC.
Note that the integrated
charm cross section in Eq. (18) includes
multiple independent parton-parton scatterings which
means
multiple charm quark pair production.
\medskip

\bigskip
\noindent{\bf 4.  Rapidity and Transverse Momentum
Distributions for Charm Production in p-p, p-Au and
Au-Au Collisions}
\medskip

In the previous section we have seen that the nuclear shadowing
function depends on the kinematic variable, $x$.  Thus, it seems
plausible that the effect that nuclear shadowing has on charm
production is not uniform throughout all of phase space.
Furthermore, the new subprocesses that are part of the
next-to-leading order calculation have different
contributions in different regions of phase space.  Thus, by
studying the single differential distributions
one might be able to get more information about the
underlying dynamics of charm production in nuclear collisions.

In general, the single
differential distribution is obtained by integrating the double
differential inclusive distribution
\begin{equation}
\frac{d N_c}{d^{2}p_{T}dy}=T_{AA}(0)
\frac{d{\sigma}_{c}
}{d^{2}p_{T}dy},
\end{equation}
where the double differential inclusive cross section
per nucleon-nucleon
interaction {\it in-medium}
is given by
\begin{equation}
\frac{d{\sigma}_{c}
}{d^{2}p_{T}dy}
=
\sum_{i,j}^{partons}
\int dx_{a} dx_{b}
\frac{F_{i}^{A}(x_{a},Q^{2})}{A} \frac{F_{j}^{A}(x_{b},Q^{2})}
{A}
\frac{d\hat{\sigma}_{i,j}
(Q^2, m_c, \hat s)}{d^{2}p_{T}dy}.
\end{equation}
The parton differential cross section calculated to
$O(\alpha_{s}^{3})$ can be written as
[2]
\begin{equation}
\frac{d\hat{\sigma}_{i,j}}{d^{2}p_{T}dy }=
\frac{\alpha^{2}_{s}}{\hat s}h^{(0)}_{i,j}+\frac{\alpha^{3}_{s}}
{2\pi \hat s^{2}}h^{(1)}_{i,j}.
\end{equation}
Expressions
for the functions $h^{(0)}_{i,j}$ and
$h^{(1)}_{i,j}$
can be found in Ref. 2.

We obtain the rapidity and transverse momentum distribution
for charm production in p-p, p-Au and Au-Au collisions by
integrating the Eq. (20) over the appropriate variable (i.e.
to obtain the rapidity distribution, for example,
we integrate Eq. (20) over the transverse
momentum).  We use two different choices of the renormalization
scale, as in the case of the total cross section, and three
different structure functions:  MRS D0, MRS D-- and
MRS A [9,10].

In Fig. 4, we present our results for the rapidity distribution
for charm production
at RHIC.  We find that different choice of the structure function
result in about
$8\%$ uncertainty at y=0, while at $y=3$ this uncertainty is
almost $40\%$.  This is due to the fact that charm production
in the large rapidity
region is probing smaller $x$-region than in case
when charm quarks are
produced in the central rapidity
region ($x_{average}\approx 0.01$ for
$y\approx 0 $ and
$x_{average}\approx 10^{-5}$ for
$y\approx 3 $).
At LHC energies,
the choice of the
structure functions introduces a much larger
theoretical
uncertainty.  From Fig. 5,  we note that
at $y\approx 3$ this uncertainty is
about a factor of $7$.  Furthermore,
we find that
the shape of the rapidity distribution
is sensitive to the
low-$x$ behavior of the structure function, resulting in a
``dip'' at $y=0$.
At $y=0$, the
$x$-values of the incoming partons
are approximately $10^{-4}$, and therefore both of the
structure functions that appear in the
distribution given by
Eq. (20) are evaluated in the region of maximum nuclear shadowing.
When $y\approx 3.5$, the
$x$-region probed is a combination of small-$x$ (maximum
shadowing) and intermediate-$x$ (small shadowing effect)
leading to the larger values for the number of
charm quarks produced than in the central rapidity region.
Furthermore, the combination of the
$x$-dependence of
nuclear shadowing effect and the steep
increase of the structure function at low-$x$
could result in less overall shadowing at
$y\approx
3.5$.  In Fig. 5 we note that
this is apparent for the MRS D-- structure function,
while the effect is much smaller
in case of MRS A and MRS D0 sets.
In Fig. 6,
 we present the results for the transverse momentum
distributions for the charm production at LHC.  We note that
the shape of $p_T$ distribution is sensitive to the
choice of the structure function.
The $p_T$ distribution is much steeper when obtained with MRS D--
structure function than with MRS A or with MRS D0 set.
The theoretical
uncertainty due to the choice of the structure function is
about order of magnitude
at low $p_T$, where the small-$x$ region is probed,
while
at larger values of $p_T$ (i.e. $p_T\geq 6GeV$)
this uncertainty is substantially reduced.
%In addition, the uncertainty due to the
%choice of the renormalization scale
%is about $30\%$ at RHIC [19] and about $40\%$ at the LHC.
In Fig. 7, we present the rapidity distribution for
charm production at LHC energies
calculated with the MRS A parton distribution and with
$Q^{2}=m_{c}^{2}$.  The two distributions without nuclear shadowing
correspond to
p-p collisions.
To obtain the number of charm quark pairs produced
 in Au-Au collisions, we need to multiply
the rapidity distribution which contains nuclear shadowing effects
(solid line) with the geometrical factor (i.e. the spatial,
overlapping function) presented in Fig. 2.
To get the number of charm quark pairs produced in central
collisions (i.e.
at zero impact
parameter), for example, one would need to multiply $d\sigma_c/dy$
by $T_{AA}(0)$.
We note that rapidity
distribution for charm production in Au-Au collisions at LHC obtained
with MRS A structure function is
flat for $\vert  y \vert \leq 3$.

The {\em effective} (in-medium)
K-factor can be defined in a way
analogous to the hadronic K-factor, namely
as the ratio of a particular
distribution for charm production in {\it nuclear} collisions
to the leading-order distribution without any nuclear
effects.  Thus, this K-factor is a measure of the size of
the effect due to higher-order corrections {\it combined}
with nuclear
effects.
In Fig. 8, we present the K-factor for the rapidity
distributions at LHC energies.  We see that over the range $-5 \leq
y \leq 5$
the  K-factor has weak dependence on $y$.
In the central rapidity region,
the K-factor for Au-Au collisions is about 1.1 (circles), while
the hadronic K-factor is 2.5 (squares).  The results for the
hadronic and for the
effective K-factor for the rapidity and transverse momentum
distributions at RHIC energies can be found in Ref. 19.

In Fig. 9 we present results for
the transverse momentum distributions at the LHC.
The calculation is done
using MRS A parton distributions and a
renormalization scale $Q^{2}=m_{c}^{2}$.  At
low-$p_T$
the nuclear shadowing effects
suppress charm production by about $62\%$,
while
the higher-order corrections, $O(\alpha_s^3$),
enhance the charm production by about
$50\%$, resulting in $p_T$ distribution which is
effectively lower than the leading-order result for
p-p
collisions\footnote{
{}From Fig. 2 we note that the nuclear shadowing of a
parton distribution in gold at $x \leq 10^{-4}$
is about $62\%$.  Therefore, combination of two parton distributions
which appears in Eq. (20) results in
about $62\%$ overall suppression
of charm production at low-$p_T$.}
At large $p_{T}$, we expect nuclear effects to become
negligible and the $p_T$ distribution to approach the
next-to-leading order results for p-p collisions.
In Fig. 10 we present the hadronic and the
{\em effective} K-factor
for the
transverse momentum distribution for charm production at
LHC energies.  We find that while hadronic K-factor
varies from $1.4$ to $7.6$
for
$0.7$ GeV$ \leq  p_{T} \leq 6$ GeV, the K-factor for Au-Au
collisions changes from $0.6$ to $4.6$ in the same $p_T$
range.
For the $x_{F}$-distribution,
we find the {\em effective} K-factor for Au-Au collisions at RHIC
(LHC) energies to increase
from $1.5$ ($1.2$) at
$x_{F}\approx0$, to around $5.2$ ($3$) at $x_{F}\approx 1$, while
the hadronic K-factor changes from $2$ ($2.4$) at
$x_{F}\approx0$, to about $6.6$ ($5$) at $x_{F}\approx 1$.

We present our results for the
single differential distribution in p-A collisions in terms
of the
parameter $\alpha_{eff}$,
defined as
\begin{equation}
\tilde{\sigma}_{AA}=A^{\alpha_{eff}}\tilde{\sigma}_{pp},
\end{equation}
where $\tilde{\sigma}_{pp}$ is a differential cross section for
charm production in hadronic collisions, and $\tilde{\sigma}_{AA}$ is
the corresponding cross section in nuclear collisions.
The parameter
$\alpha_{eff}$ is
a sum of the geometrical contribution and
the nuclear shadowing effect
(i.e. $\alpha_{eff}=\alpha + \alpha_{NS}$, $\alpha_{NS} < 0$).
The later effect reduces the value of the
``standard'' $\alpha$, which
for minimum bias A-A
collisions is $2$, while for the central A-A collisions,
$\alpha=4/3$.  The values for
$\alpha_{NS}\equiv
\alpha_{eff}-\alpha$
in p-A and A-A collisions at RHIC and LHC
are presented in Fig. 11.
Knowing $\alpha_{eff}(y)$ and
$\alpha_{eff}(p_T)$
it is possible to extract
the
rapidity and the transverse momentum
distribution for p-A collisions
by suitable modification of the distribution for charm production
in p-p
collisions.  For example,
$\frac {d\sigma_c^{pA}}{dy}=
A^{\alpha_{eff}(y)}
\frac {d\sigma_c^{pp}}{dy}$.  The K-factor for inclusive
distribution for charm production
in p-A collisions can also be determined from
$\alpha_{eff}$,  as
$K=(\frac {d\sigma_c^{pA}}{dy})/
{(\frac {d\sigma_c^{pp}}{dy})}_{LO}$.

\medskip
\noindent{\bf 5.  Conclusions}
\medskip

To summarize, we have presented results for the
differential and total inclusive cross sections for
charm production in hadronic and heavy-ion collisions at
RHIC and LHC energies.  We have included the next-to-leading
order corrections, $O(\alpha_s^3)$, and the nuclear
shadowing effect with the assumption that the shadowing effect is
the same for the gluon density in a nucleus as the observed effect
in the quark distribution [16].
We have shown that low-energy
data for the total cross section for charm production in
p-p and p-A collisions
provide some constraints on out theoretical parameters,
especially for the choice of the renormalization scale.
The choice $Q=m_c$ seems to be preferred by the data.
Theoretical uncertainty in the calculation of
charm production in nuclear collisions due to the choice of
the structure function is small at Fermilab fixed target
energies (only few percent) and at RHIC ($8\%$
in the central rapidity region
), while at LHC
energies this uncertainty is about a factor of $6$.
Furthermore, the shape of the rapidity distribution for
charm production in Au-Au collisions
at LHC
is very sensitive to the low-$x$ behavior of the gluon structure
function, resulting in a larger ``dip" at $y=0$
for a more singular function.
Similarly,
the transverse momentum distribution at LHC is
much steeper when obtained with
MRS D-- structure function than with less singular structure
function, such as
MRS A set or the non-singular structure function, such as
MRS D0 set.
For the rapidity distribution,
we have found the hadronic K-factor to be
about $2$ ($2.5$) at RHIC (LHC) energies, while the
{\it effective} K-factor for Au-Au collisions
is about $1.5$ ($1.1$).
In case of $p_T$ distribution, the
hadronic K-factor at LHC energies
varies from
$1.4$ at $p_T=0.7$ GeV to $7.6$ at $p_T=6$ GeV,
while the {\it effective} K-factor changes from
$0.6$ to $4.6$ in the same $p_T$ range.  This behavior of
the {\it effective} K-factor is a direct consequence of the
fact that the low $p_T$ region (or small $x$) corresponds to
the maximum shadowing of the gluon distribution, which in
case of gold is about $62\%$, while the larger values of
$p_T$ probe region of phase space where the nuclear shadowing
is smaller.
We have also obtained the effective A-dependence for the differential and
total cross sections
for charm production in p-Au collisions
at RHIC and in Au-Au collisions at the LHC.
We have found that the dominant contribution
to charm production
comes from the initial state gluons
(about $95\%$ at RHIC and
$99\%$ at LHC).
Thus,
combined measurements of inclusive charm production in hadronic and
nuclear collisions at these energies,
in addition to providing an important test of perturbative QCD in
the small-$Q^2$ and small-$x$ region,
might be able to
provide
valuable information about the
elusive role of gluons
inside a nucleus, especially
in the region of very small $x$.
\vfil\eject

\centerline{\bf Acknowledgements}
\vskip 0.2true in
\par
We would like to thank M. Mangano and P. Nason
for providing us with
the fortran routines for calculating double differential
distributions for charm production in hadronic collisions.
This work was supported in part through
U.S. Department of Energy Grants Nos.
DE-FG03-93ER40792 and
DE-FG02-85ER40213.

\bigskip

\centerline {\bf References}
\vskip 0.2true in
\noindent
[1]  P. Nason, S. Dawson
and R. K. Ellis, Nucl. Phys. {\bf B327} (1989) 49; ibid.
Nucl. Phys. {\bf B303} (1988) 607.
\vskip 0.15true in
\noindent
[2]  M. L. Mangano, P. Nason and G. Ridolfi, Nucl. Phys.
{\bf B373} (1992) 295.
\vskip 0.15true in
\noindent
[3] For a review of the future plans for
high precision charm measurements
see, {\em Proceedings of the
Workshop on the Future of High Sensitivity Charm Experiments:
CHARM2000, Batavia, IL. 7-9 June 1994}.
\vskip 0.15true in
\noindent
[4] E. Shuryak, Phys. Rev. Lett. {\bf
68}
(1992) 3270; K. Geiger,
Phys. Rev. {\bf D48} (1993) 4129.
\vskip 0.15true in
\noindent
[5]  For example, see
B.  Muller, in {\it Particle Production in Highly Excited Matter}, eds.
H. Gutbrod and J. Rafelski (Plenum, New York, 1993).
\vskip 0.15true in
\noindent
[6]  For a recent review of the thermal photons and dileptons see,
V. Ruuskanen, in {\it Proceedings of Quark Matter '90},
Nucl. Phys. {\bf A525} (1991) 255c and on
$J/\psi$ suppression see, S.  Gavin,
in {\it Proceedings of the
Second International Conference on Physics and Astrophysics
of Quark Gluon Plasma}, Calcutta, 19-23 January 1993, in press.
\vskip 0.15true in
\noindent
[7] A. Bohr and B. R. Mottelson,
Nuclear Structure I (Benjamin, New
York, 1969) pp. 160.
\vskip 0.15true in
\noindent
[8]
For a recent review of the structure functions and HERA
physics see, G. Wolf, DESY preprint, DESY 94-022.
\vskip 0.15true in
\noindent
[9]
A. D. Martin, W. J. Sterling and R. G. Roberts, Phys. Rev.
{\bf D47} (1993) 867.
\vskip 0.15true in
\noindent
[10]
A. D. Martin, W. J. Sterling and R. G. Roberts,
RAL preprint, RAL-94-055.
\vskip 0.15true in
\noindent
[11]  S. Aoki {\it et al.}, Phys. Lett. {\bf B224} (1989) 441;
S. P. K. Tavernier, Rep. Prog. Phys. {\bf 50} (1987) 1439;
S. Barlag {\it et al.}, Z. Phys. {\bf C39} (1988) 451.
\vskip 0.15true in
\noindent
[12]  For a recent review of the factorization theorems in
perturbative QCD see, J. C. Collins, D. E. Soper and G. Sterman, in
{\it Perturbative Quantum Chromodynamics}, ed. A. H. Mueller
(World Scientific, Singapore, 1989).
\vskip 0.15 true in
\noindent
[13] K. J. Eskola, K. Kajantie
and J. Lindfors, Nucl. Phys. {\bf B323}
(1989) 37.
\vskip 0.15true in
\noindent
[14] L. Durand and H. Pi, Phys. Rev. D{\bf 38}, 78 (1988).
\vskip 0.15true in
\noindent
[15]
J. Qiu, Nucl. Phys. {\bf B291} (1987) 746; K. J. Eskola,
Nucl. Phys. {\bf B400} (1993)  240.
\vskip 0.15true in
\noindent
[16]
NMC Collaboration, P. Amaudruz {\it et al.},
Z. Phys. {\bf C51} (1991) 387;
E665 Collaboration, M. R. Adams {\it et al.},
Phys. Rev. Lett. {\bf {68}} (1992) 3266; EMC Collaboration,
J. Ashman
{\it et al.}, Phys. Lett. {\bf B202} (1988) 603;
M. Arnedo {et al.}, Phys. Lett. {\bf B211} (1988) 493.
\vskip 0.15true in
\noindent
[17]
C. J. Benesh, J. Qiu and J. P Vary,  Los Alamos preprint,
LA-UR-94-784.
\vskip 0.15true in
\noindent
[18] K. J. Eskola, J.-W. Qiu, and X.-N. Wang,
Phys. Rev. Lett. {\bf 72}, 36 (1994).
\vskip 0.15true in
\noindent
[19]  I. Sarcevic and P. Valerio, University of Arizona preprint
AZPH-TH/93-13, to be published in Phys. Lett. {\bf B} (1994).
\vskip 0.15true in
\noindent

%[1]  For a recent review see, {\it Quark Matter '93,
%Proceedings of the Tenth International
%Conference on Ultrarelativistic Nucleus-Nucleus Collisions},
%ed. H. A. Gustaffson,
%Nucl.\ Phys.\ {\bf A566} (1994).
%\vskip 0.15true in
%\noindent
%[7]  For a recent review of the factorization theorems in
%perturbative QCD see, J. C. Collins, D. E. Soper and G. Sterman, in
%{\it Perturbative Quantum Chromodynamics}, ed. A. H. Mueller
%(World Scientific, Singapore, 1989)
%\vskip 0.15 true in
%[18]  D. Kharzeev and H. Satz, Phys. Lett, {\bf B327}, 361 (1994);
%A. Bialas and Czyz,
%Phys. Lett. {\bf B328}, 172 (1994).
\vfil\eject
\begin{center}
{\bf Figure Captions}
\end{center}
Fig. 1. The total cross section for charm quark production in
proton-proton collisions calculated to the next-to-leading order
(LO+NLO) for values of $E_{beam}$
ranging from $50$GeV to $2$TeV
%The solid lines represent our calculations
for two
renormalization scales, $Q=m_{c}$ and $Q=2m_{c}$.
The results
are compared to the data from  p-p and p-A
collisions [11].
\vskip 0.15true in
Fig. 2. The nuclear overlapping functions, $T_{AB}(b)$ for p-Au and
Au-Au collisions obtained with two different
nuclear density functions,
a)  Woods-Saxon [7] and b) the Gaussian form.
\vskip 0.15true in
Fig. 3.
The nuclear shadowing functions
given by Eq. (17) (solid line) and
by Eq. (16)
(dashed line) fitted to the
EMC, NMC and E665 deep-inelastic
lepton-nucleus data [16].  We include a plot of the nuclear
shadowing
function for Au given by Eq. (17) (solid line), which is used in
our calculation of charm production in p-Au and Au-Au
collisions.
\vskip 0.15true in
Fig. 4.  The rapidity distribution for charm quark production in Au-Au
collisions at RHIC energies
calculated to $O(\alpha_s^3)$ and with three different sets of
structure functions:
MRS A (solid line),
MRS D--
(dotted line),
and MRS D0 (dashed line).
The renormalization scale is
taken to be $Q=m_c$.
\vskip 0.15true in
Fig. 5.
Same
as Fig 4. but for LHC.  The curves
are labeled as in Fig. 4.
\vskip 0.15true in
Fig. 6.
Same
as Fig. 5, but for the transverse  momentum
distribution.  The curves are labeled as in Fig. 5.
\vskip 0.15true in
Fig. 7.  The rapidity distribution for charm quark production in
proton-proton and Au-Au collisions at LHC energies
calculated to the next-to-leading order (LO+NLO) with
nuclear shadowing effects (NS) (solid line),
without NS (i.e. for p-p collisions) (dotted line),
only leading-order (LO) (dashed line) and LO without NS
(long-dashed line).
\vskip 0.15true in
Fig. 8.  The {\em effective} (in-medium) K-factor, defined as
$K\equiv(\frac{d\sigma_{c}^{AA}}{dy})/
(\frac{d\sigma_{c}^{pp}}{dy})_{LO}$,
for
the distributions in Fig. 7.  The K-factor for p-p collisions
is about
$2.5$ (squares), while the
K-factor {\it in-medium}
(in case of Au-Au collisions) is about $1.1$ (circles) in the
central rapidity region.
\vskip 0.15true in
Fig. 9.  Same as Fig. 7, but for
the transverse
momentum distribution.
Curves are labeled as in Fig. 7.
\vskip 0.15true in
Fig. 10. The same as Fig.  8 but for
the transverse momentum
distributions presented in Fig. 9.  The K-factor for p-p
collisions ranges from $1.4$ at $p_{T}=0.7$ GeV to $7.6$
at $p_{T}=6$ GeV.  In the same $p_{T}$ range, the
K-factor {\it in-medium} increases from $0.6$ to $4.6$.
\vskip 0.15true in
Fig. 11. The plot of $\alpha_{eff}-\alpha$ as a function of
$y$ and
$p_{T}$ for p-Au and Au-Au collisions.
The $\alpha_{eff}$ is defined as
$\tilde{\sigma}_{Au-Au/p-Au}=A^{\alpha_{eff}}
\tilde{\sigma}_{p-p}$,
where $\tilde{\sigma}_{p-p}$ is a differential
cross section for charm
production in p-p collisions and
$\alpha_{eff}=\alpha + \alpha_{NS}$.  The
parameter $\alpha$, frequently used in the literature,
corresponds to
the A-dependence coming
from the geometry only.  In the limit of no
nuclear shadowing,
$\alpha_{eff}\rightarrow \alpha$.  In case of central A-A collisions,
$\alpha=4/3$, while for minimum bias collisions
$\alpha=2$.  For p-A minimum bias collisions, $\alpha=1$.
\end{document}